\documentclass[]{spie}  

\usepackage{amsmath,amsfonts,amssymb}
\usepackage{graphicx}
\usepackage[colorlinks=true, allcolors=blue]{hyperref}
\usepackage{float}

\title{The FastrSHWFS Development \& Test Results}

\author[1]{Natasha Popenoe}
\author[1]{Benjamin Gerard}
\author[1]{Dominic Sanchez}
\affil[1]{Lawrence Livermore National Laboratory, \textit{ 7000 East Ave Livermore CA USA}}

\authorinfo{Further author information: (Send correspondence to N.P.)\\N.P.: E-mail: popenoe1@llnl.gov, Telephone: 1 301 204 3629\\  B.G.: E-mail: gerard3@llnl.gov, Telephone: 1 925 423 3957}

\begin{document} 
\maketitle

\vspace{3cm}

\begin{abstract}
The recent 2020 Decadal Survey of Astronomy and Astrophysics listed habitable
exoplanet imaging with future extreme adaptive optics (AO) on 30m-class telescopes as
a key priority in the coming decade. However, there is a current 100x contrast gap
between the best systems today and what is needed to enable this goal. Astronomical
AO is a required approach to enable ground-based diffraction-limited imaging of
exoplanets on future extremely large telescopes. Time lag between the end of an
exposure and the application of deformable mirror commands is a major contributor to
the error budget in many AO systems, and detector read time is often a large component
of this lag. We present two designs for a modified Shack Hartmann wavefront sensor
(SHWFS), named Focal plane Actualized Shifted Technique Realized for a SHWFS
(fastrSHWFS), to reduce the time lag component.  This design steers the spot pattern at
the focal plane into a rectangular or linear array with a custom aspect ratio, reducing
readout time. These masks were made by a recently available gray-scale two photon polymerization 3D printing technology that in principle provides sufficient depth resolution and dynamic range for such masks. The mask with focus yields aberrated results while the mask with tip/tilt
only yields some defined spots. This manuscript outlines the current SHWFS concept, our
fastrSHWFS theoretical solution to addressing time lag, reflection and quality analysis
of printed mask designs, and results from testing both masks on the High Contrast
Testbed at Lawrence Livermore National Lab. This work follows the test of a previous fabricated
fastrSHWFS masks that had insufficient optical quality.
\end{abstract}

\keywords{Adaptive optics, Exoplanet direct imaging, Wavefront sensing, Bandwidth error}

\section{INTRODUCTION}
Direct imaging of exoplanets provides snapshots of planetary systems outside our solar system, revealing crucial details about exoplanet physical characteristics, atmospheric composition, orbital dynamics, and potential for life, offering unique insights into planet formation processes and stellar system evolution. Imaging exoplanets from ground-based observatories is challenging due to Earth’s atmosphere causing aberrations in the light collected from far away sources. In Figure 1, we can see that direct imaging makes up a small portion of total (about 6000) detected exoplanets, and these directly imaged detections are largely at longer orbital periods (farther from their host star) and have large mass. Our current exoplanet imaging techniques that allow us to image planets about 10AU from their host stars. However, terrestrial planets around Sun-like stars that have the possibility of hosting life are around 1-2AU away from their star, so improvements to our current techniques are necessary for possibly discovering life [7].
\label{sec:intro}  
   \begin{figure} [H]
   \begin{center}
   \begin{tabular}{c} 
   \includegraphics[height=8cm]{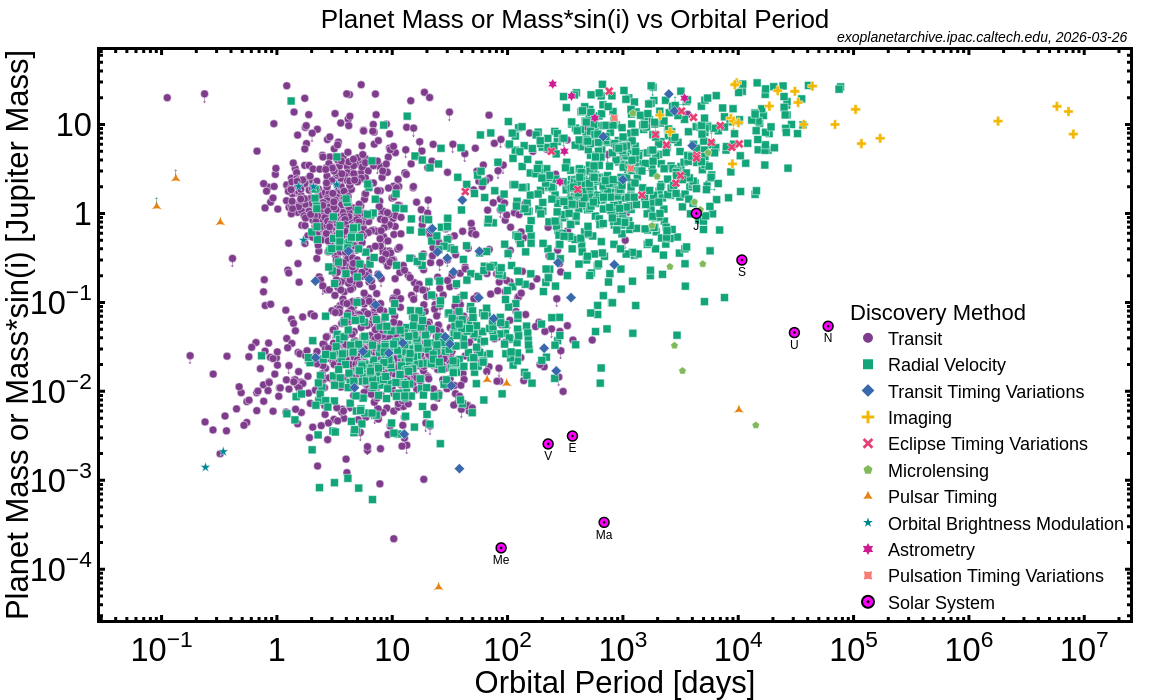}
   \end{tabular}
   \end{center}
   \caption[example] 
   { \label{fig:example} 
Total number of exoplanet detections split up by their detection method and plotted by orbital period and mass. Directly imaged detections are shown by the yellow plus symbol [5].}
   \end{figure} 

\section{BACKGROUND}
\subsection{Shack-Hartmann Wavefront Sensor}
AO is a way to correct for Earth’s atmospheric aberrations that limit our ability to directly image planets from the ground. AO works using a wavefront sensor (WFS), a real-time controller (RTC), and a deformable mirror (DM). A Shack Hartmann Wavefront Sensor (SHWFS) is the standard WFS used in most AO systems. The process is as follows: A flat wavefront enters a telescope pupil and intersects an array of lenslets, all focusing light to a focal plane yielding a grid of spots shown in Figure 2. If a distorted wavefront enters the telescope pupil and intersects the same array of lenslets, each lens will focus its light at a slightly shifted position from its center point, causing a grid of displaced spots. 

   \label{sec:intro}  
   \begin{figure} [h]
   \begin{center}
   \begin{tabular}{c} 
   \includegraphics[height=6cm]{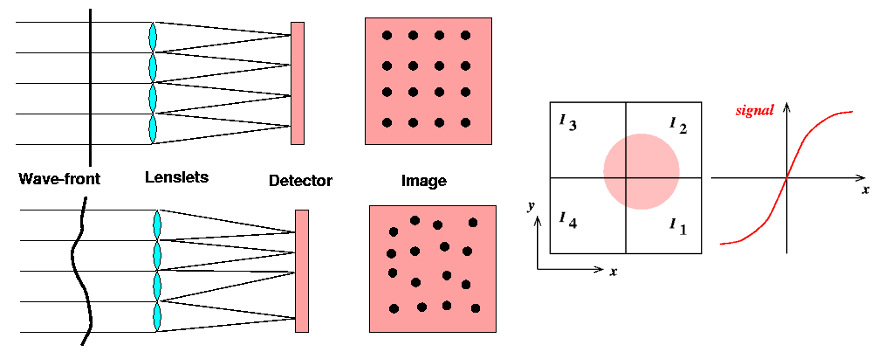}
   \end{tabular}
   \end{center}
   \caption[example] 
   { \label{fig:example} 
SHWFS lenslet array for a flat, un-aberrated wavefront (upper left) and distorted wavefront (lower left). A quad-cell layout of spot displacement calculation (right) [1]}
   \end{figure}
   
The size of the spot on the detector is determined by the diffraction limit, composed of the diameter of the lenslet (d) and distance from lenslet array to detector (L) shown in Eqn. 1.
\begin{equation}
h_{\text{diffraction}} = 1.22\, L \frac{\lambda}{d}
\end{equation}

If spots are within the diffraction limit, they will be well defined in their quad-cell. The displacement of each of those spots is proportional to the local slope of the wavefront at that point, calculated by the RTC. That information is used to reconstruct the wavefront at the telescope pupil and is deconstructed into modes. These modes are issued as a command to the deformable mirror to correct for that atmospheric aberration on the next iteration of the adaptive optics loop shown in Figure 3. In a best-case scenario, this AO loop will move quickly enough that by the time the DM pistons are actuated to flatten out an incoming wavefront, the atmosphere has not changed dramatically. This AO loop makes the stellar point spread function (PSF) stable enough to detect faint exoplanets nearby the host star despite atmospheric turbulence. 

   \label{sec:intro}  
   \begin{figure} [h]
   \begin{center}
   \begin{tabular}{c} 
   \includegraphics[height=6cm]{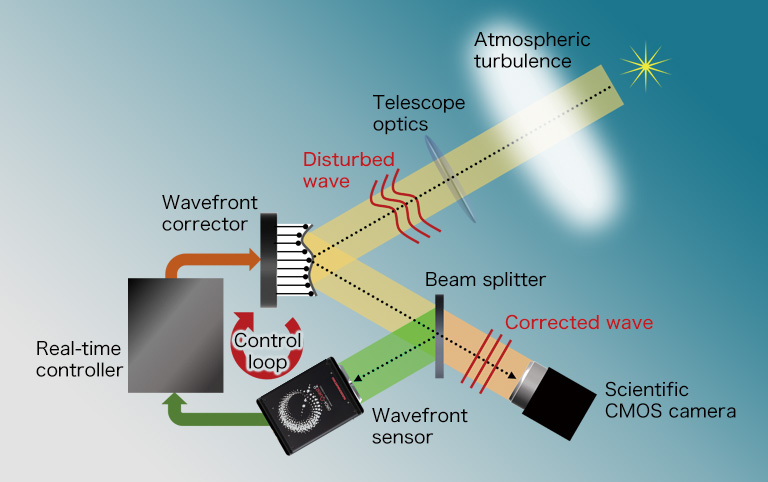}
   \end{tabular}
   \end{center}
   \caption[example] 
   { \label{fig:example} 
An adaptive optics control loop. The ‘wavefront corrector’ is the deformable mirror in the standard case [6]}
   \end{figure}

\subsection{Bandwidth Error}
The Gemini Planet Imager (GPI) is a next generation adaptive optics instrument built with the goal of imaging exoplanets. Upon evaluation of the GPI AO performance, the results conclude that for all targets and atmospheric conditions, AO bandwidth error is the dominant error term in many extreme AO systems [8]. Within the total latency, the detector read time is a large contributor.

\begin{equation}
\sigma_{\text{bandwidth}}^2 \propto (f_g \, \tau)^{5/3}
\tag{2}
\end{equation}

\begin{equation}
\tau = \frac{1}{2} \, \text{exposure} + \text{readout} + \text{transfer} + \text{processing} + \text{DM}_{\text{drivers}} + \text{DM}_{\text{electronics}} + \text{DM}_{\text{physical}}
\tag{3}
\end{equation}
Bandwidth error (Eqn. 2), is a term proportional to the Greenwood frequency, a variable that represents how fast the atmosphere changes, times the latency to the 5/3 power. The latency in an AO system (Eqn. 3), is an addition of the time it takes for each component of the AO loop to complete [9]. 
   \label{sec:intro}  
   \begin{figure} [H]
   \begin{center}
   \begin{tabular}{c} 
   \includegraphics[height=8cm]{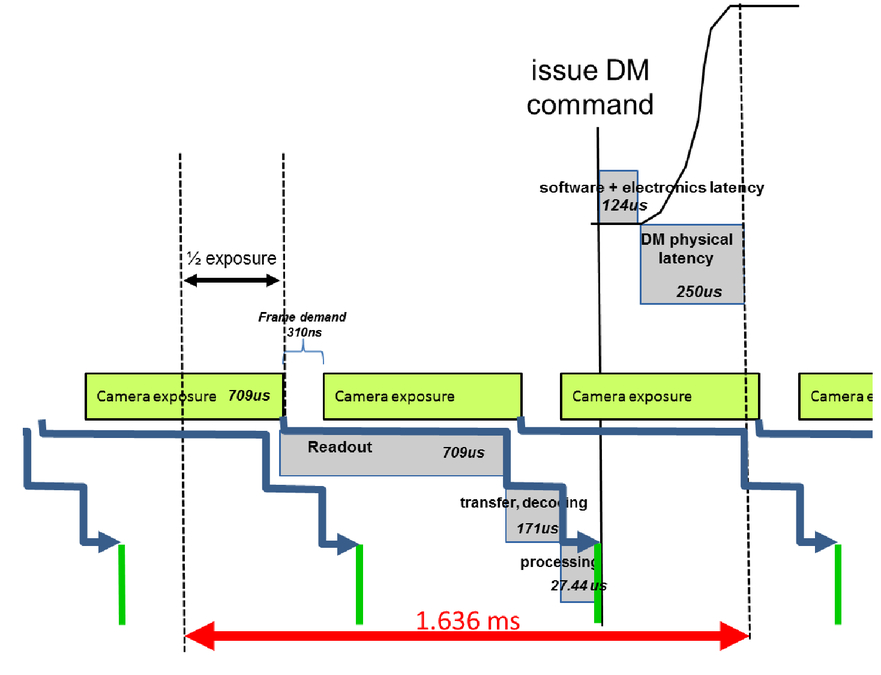}
   \end{tabular}
   \end{center}
   \caption[example] 
   { \label{fig:example} 
Every microsecond delay in AO latency from camera exposure (green) to when the DM actuator moves shown in grey. Readout is the largest contributor in this example.
The readout contributes to almost half of the total latency.}
   \end{figure}

AO bandwidth error is correlated with causing a wispy ‘butterfly pattern’ surrounding stellar PSFs which limit our ability to image exoplanets near to their host star [8]. Terrestrial planets around Sun-like stars that have the possibility of hosting life are around 1-2AU away from their host star [7]. Our current exoplanet imaging techniques allow us to image planets about 10AU from their host stars according to Figure 1.
 \begin{figure} [H]
   \begin{center}
   \begin{tabular}{c} 
   \includegraphics[height=5cm]{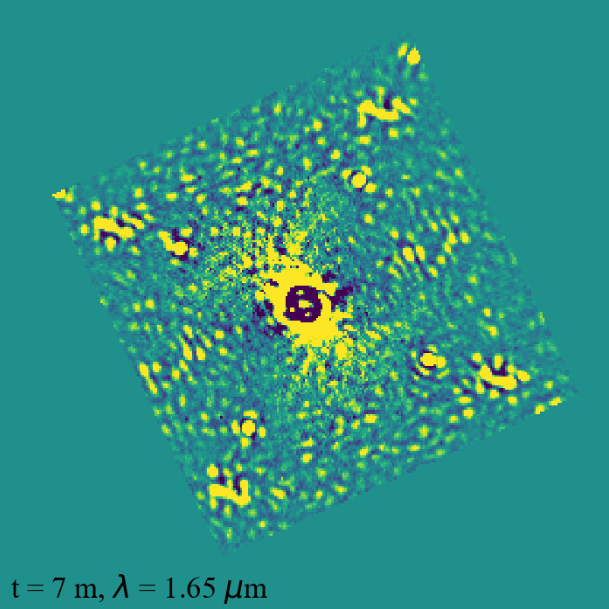}
   \includegraphics[height=5cm]{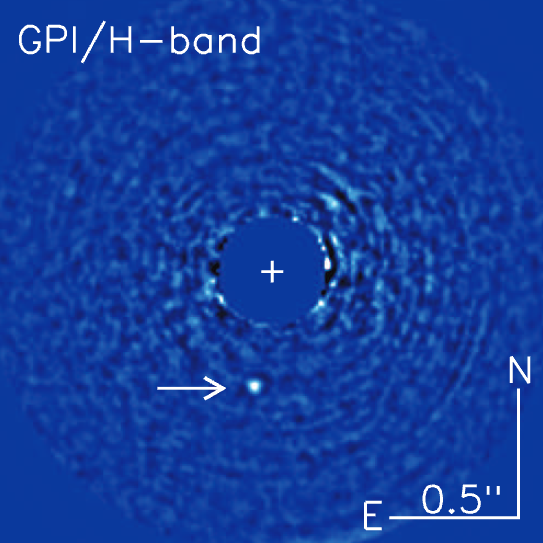}
   \end{tabular}
   \end{center}
   \caption[example] 
   { \label{fig:example} 
AO bandwidth error causes ‘butterfly effect’ limiting contrast at small inner working angles of unprocessed image of exoplanet system 51 Eri (left) [3]. Processed image of exoplanet 51 Eri b (right) [4].}
   \end{figure} 
 The butterfly effect surrounding 51 Eri caused by AO latency is shown in Figure 5. There are a few strategies\footnote{ to reduce bandwidth error: (1) run the AO system faster (reducing the half-frame delay), (2) reduce the computational latency with faster computers and/or algorithms, (3) implement predictive control algorithms, and/or (4) reduce the hardware latency of the wavefront sensor camera and/or deformable mirror [2].} to reduce the latency that prevents us from detecting closely separated exoplanets. This paper outlines the approach that reduces the hardware latency of the wavefront sensor camera. The fastrSHWFS changes the wavefront sensor spot pattern so that it occupies fewer rows of the detector compared to the SHWFS, reducing the read time and the total system latency.

\section{MODELING RESULTS}
\subsection{Design}

Compared to the SHWFS circular array of spots, the fastrSHWFS steers beams into a line (at the most extreme case) or rectangular array of spots shown in Figure 6. This design yields fewer rows on the detector than the SHWFS which reduces detector readout time because, in general, optical/near infrared wavefront sensor cameras read out voltages in rows. 
\begin{figure} [H]
   \begin{center}
   \begin{tabular}{c} 
   \includegraphics[height=5cm]{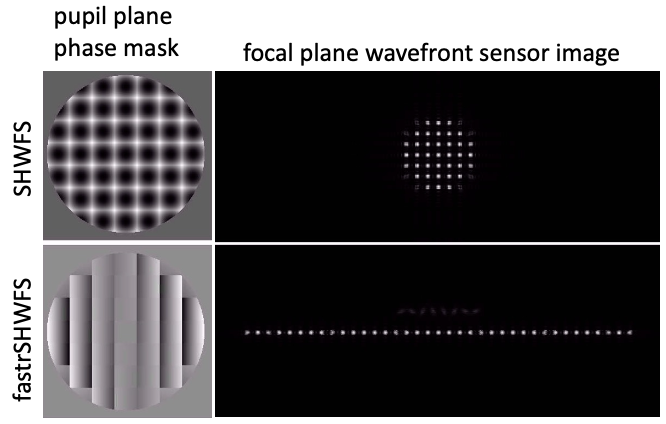}
   \end{tabular}
   \end{center}
   \caption[example] 
   { \label{fig:example} 
FastrSHWFS simulated concept vs standard SHWFS [2].}
   \end{figure} 

\begin{figure} [H]
   \begin{center}
   \begin{tabular}{c} 
   \includegraphics[height=6.7cm]{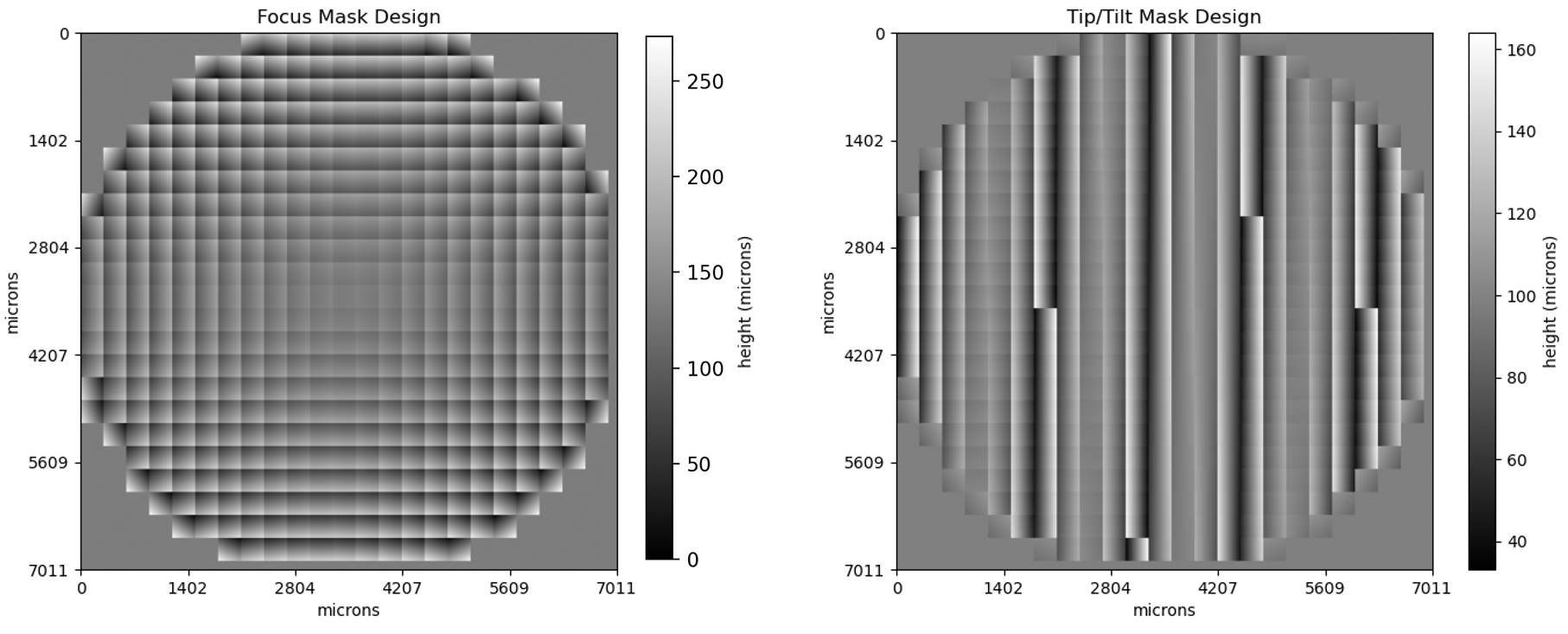}
   \end{tabular}
   \end{center}
   \caption[example] 
   { \label{fig:example} 
Height profiles in microns for focus mask design (left) and tip/tilt only design (right). }
   \end{figure} 
Two designs, shown in Figure 7, were fabricated using Nanoscribe's new Qx machine, which uses a variable voxel size to enable down to 10 nm step sizes with up to about 650 microns of print depth dynamic range and removes stitching errors, thus seeming appealing to print sufficient quality fastrSHWFS masks. Both design are reflective masks, one with focus implemented into each mirrorlet and one with tip tilt only which requires a focusing lens downstream. The focus mask distributes spots into a line that spans the width of the detector at the focal plane. The tip tilt mask redistributes spots into a rectangular array at the focal plane.
 
\subsection{Reflection Geometry}
\subsubsection{Focus Mask}
In Python, we modeled the reflection geometry of both masks to visualize the beam width at every z position downstream of the masks and examine when the spots came into focus \footnote{To see these played in movie form or with a slider go to https://github.com/npopenoe/fastrSHWFS.git}. Each mirrorlet subaperture for the focus mask is rounded to focus incoming light downstream. In Python, we fitted a plane to each of these mirrorlet subapertures and calculated the reflection geometry using the normal of the surface of each of those planes. Those normals were projected downstream in the z direction and with a movie (snapshots shown in Figure 8), we confirmed that the simulated and designed results matched with a focal plane 2.7mm downstream of the focus mask. 

 \begin{figure} [H]
   \begin{center}
   \begin{tabular}{c} 
   \includegraphics[height=14cm]{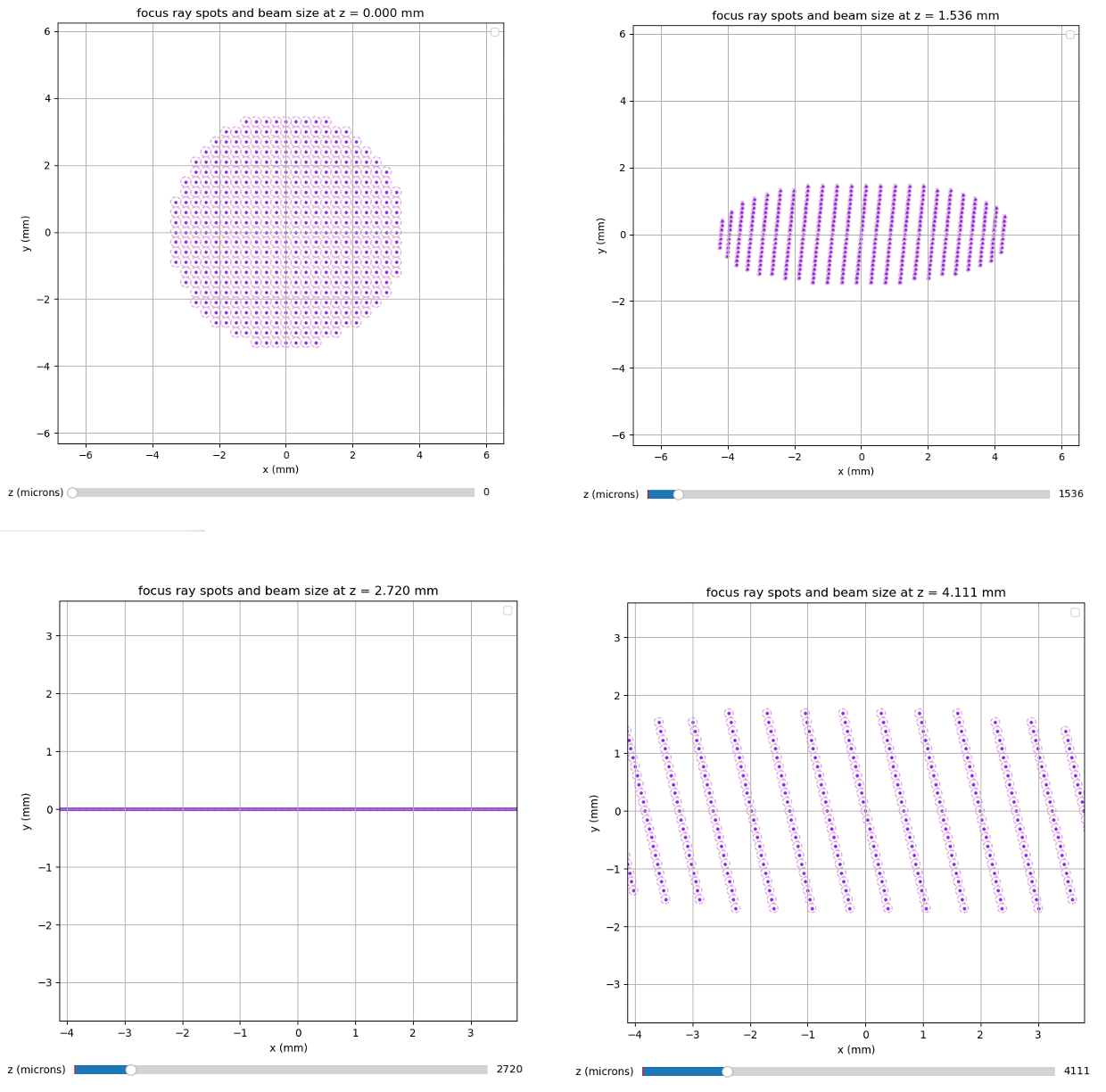}
   \end{tabular}
   \end{center}
   \caption[example] 
   { \label{fig:example} 
Focus mask snapshots simulation downstream of mask. Spots at z=0(top left). Spots beginning to converge (top right). Spots converge to focus (bottom left). Spots then diverging as we move out of the focal plane (bottom right). Axes in mm and slider in microns. }
   \end{figure} 

\subsubsection{Tip/tilt Mask}
The tip tilt mask analysis is simpler considering there is no focus implemented to the mirrorlets, so each subaperture is already flat. A projection of the normal at each surface showed that the beams diverged outward as expected. If a “lens” is placed downstream of the mask, meaning the geometry of the ray itself is altered in Python to mimic a paraxial lens placed at some z after reflection, these spots come into focus into a rectangular array shown in Figure 9. The orientation of these spots was confirmed by the expected design, 4 to 5 spots for the height of the rectangle and 100 spots for the width. 

 \begin{figure} [H]
   \begin{center}
   \begin{tabular}{c} 
   \includegraphics[height=11cm]{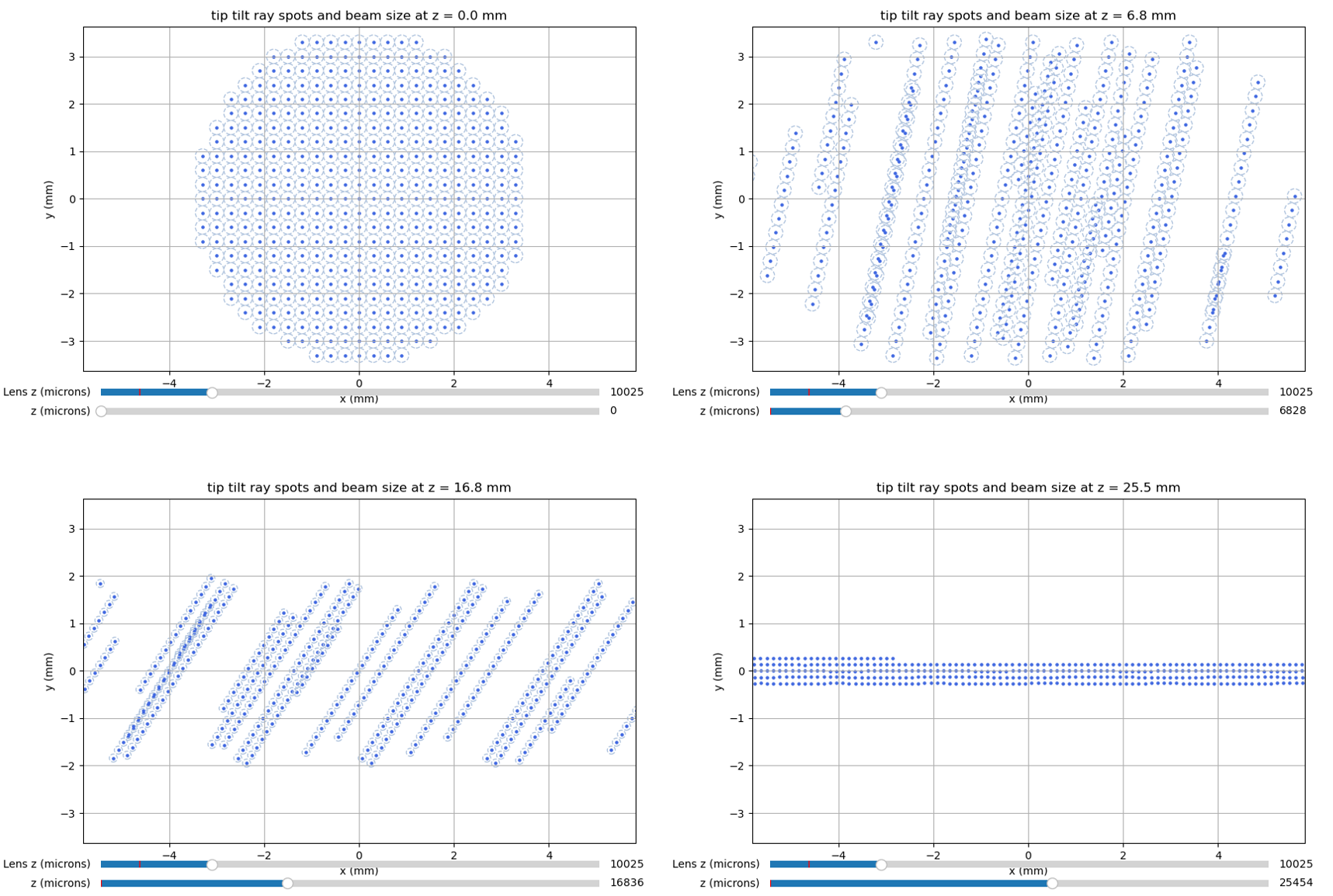}
   \end{tabular}
   \end{center}
   \caption[example] 
   { \label{fig:example} 
Tip tilt mask snapshots simulation downstream of mask with lens at z=10mm. Spots at z=0 (top left). Spots diverge outward (top right). Spots begin to converge after lens (bottom left). Spots converge to focus (bottom right). Axes in mm and sliders in microns.}
   \end{figure} 

\subsection{Laboratory Layout}
Most optical setups follow a telecentric lens system which preserves object magnification throughout the system producing images without added distortions and ensures light rays are parallel to the optical axis. For our system, this would require below f/1 lenses downstream of the mask to be placed at a distance from the spot plane equal to its focal length. The f number equal to 1 indicates that the lens's aperture size D is equal to its focal length F shown in Eqn. 4. 

\begin{equation}
\ {f}\# = \frac{F}{D} \
\tag{4}
\end{equation}

Plotting reflected beam width versus distance from both masks, we can see for the focus mask, to achieve a telecentric system we would need a lens of f number less than 1 (which is impossible to manufacture) shown in Figure 10. For the tip/tilt mask, the system reaches f/1 $\sim 45mm$ away from the mask shown in the same figure. Referencing the reflected beam width versus distance from the mask plot in Figure 11, we can see that for the tip/tilt mask, the beam width is almost 2 inches wide, meaning we’d require a 2-inch lens to be placed $\sim 45mm$ away from the mask to capture all the light and preserve telecentricity. 

 \begin{figure} [ht]
   \begin{center}
   \begin{tabular}{c} 
   \includegraphics[height=7cm]{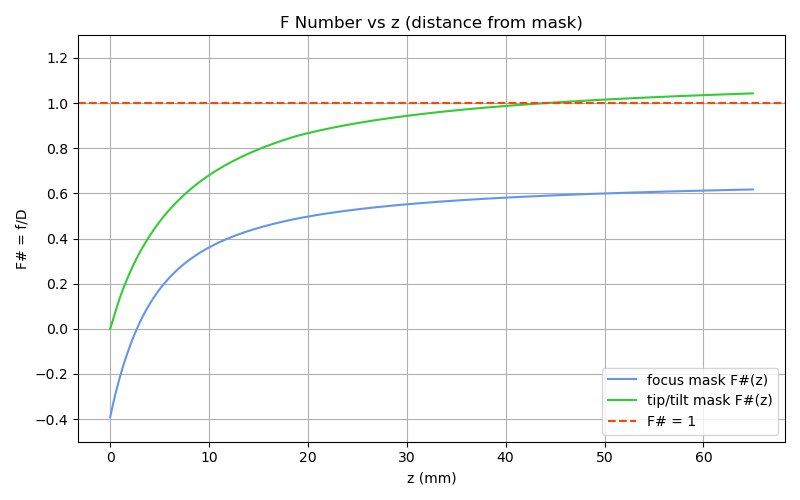}
   \end{tabular}
   \end{center}
   \caption[example] 
   { \label{fig:example} 
The f number (y axis) versus distance in mm from mask (x axis) for both focus mask in blue and tip/tilt in green. f/1 marked with dashed red. }
   \end{figure} 

 \begin{figure} [ht]
   \begin{center}
   \begin{tabular}{c} 
   \includegraphics[height=7cm]{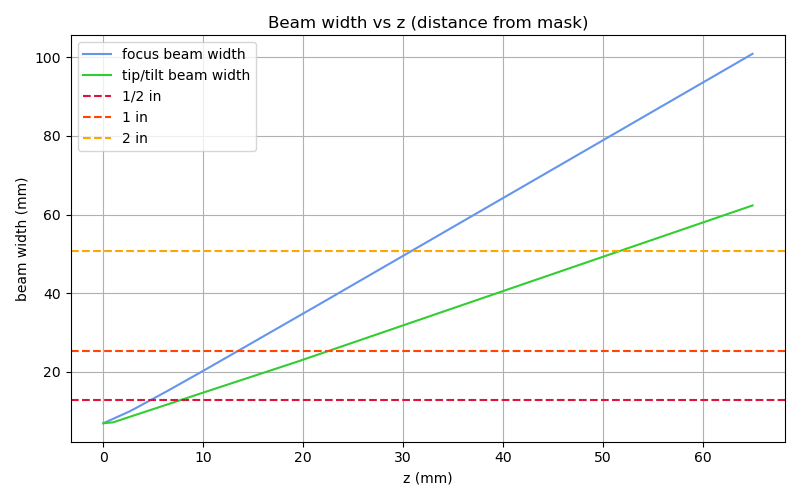}
   \end{tabular}
   \end{center}
   \caption[example] 
   { \label{fig:example} 
Beam width (y axis) versus distance from mask (x axis) for both focus mask in blue and tip/tilt in green. Beam size marked with dashed lines for various lens sizes.}
   \end{figure} 

Downstream we are focused on capturing as much light as possible (having a lens size larger than the beam width). We are not as concerned with preserving the pupil, so having a telecentric system is not necessarily required for the mask with focus. Our laboratory layout, shown in Figure 12 and 13, has downstream lenses placed very close to the mask at non-telecentric distances. We designed this laboratory layout according to the available space on the testbed and used majority of lenses already in stock at the lab (no custom optics other than the fastrSHWFS masks). 

 \begin{figure} [ht]
   \begin{center}
   \begin{tabular}{c} 
   \includegraphics[height=6cm]{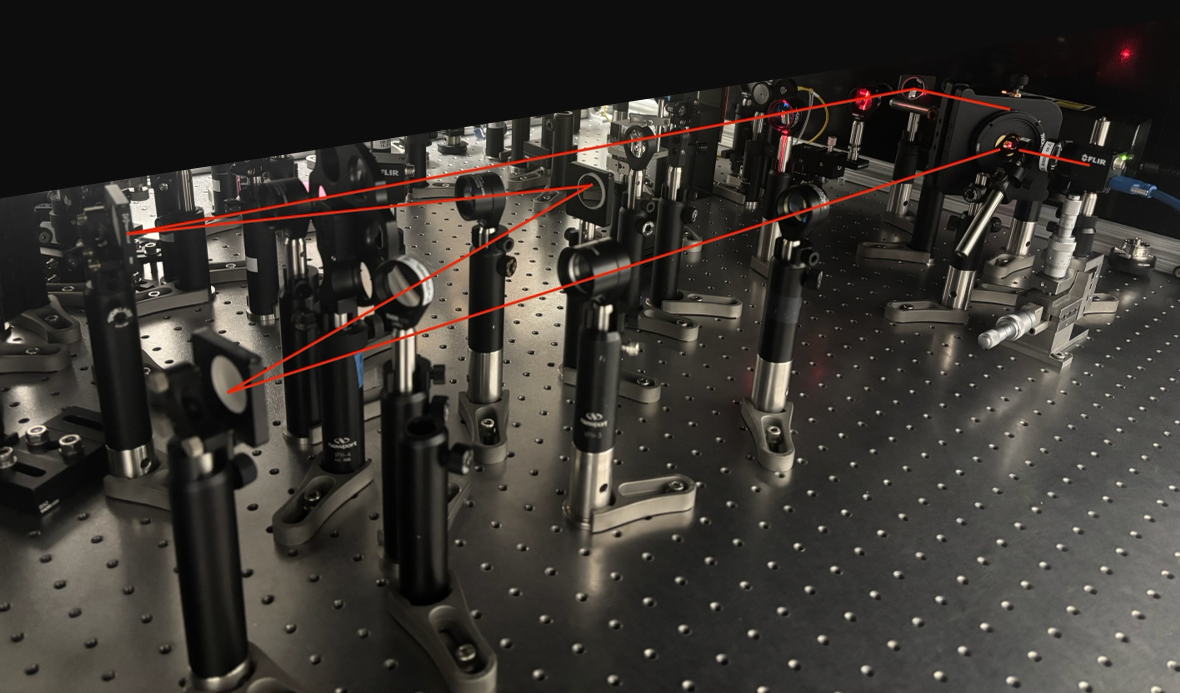}
   \end{tabular}
   \end{center}
   \caption[example] 
   { \label{fig:example} 
Photograph of the LLNL High Contrast Testbed (HCT) lab setup for mask with focus. }
   \end{figure} 

 \begin{figure} [ht]
   \begin{center}
   \begin{tabular}{c} 
   \includegraphics[height=6cm]{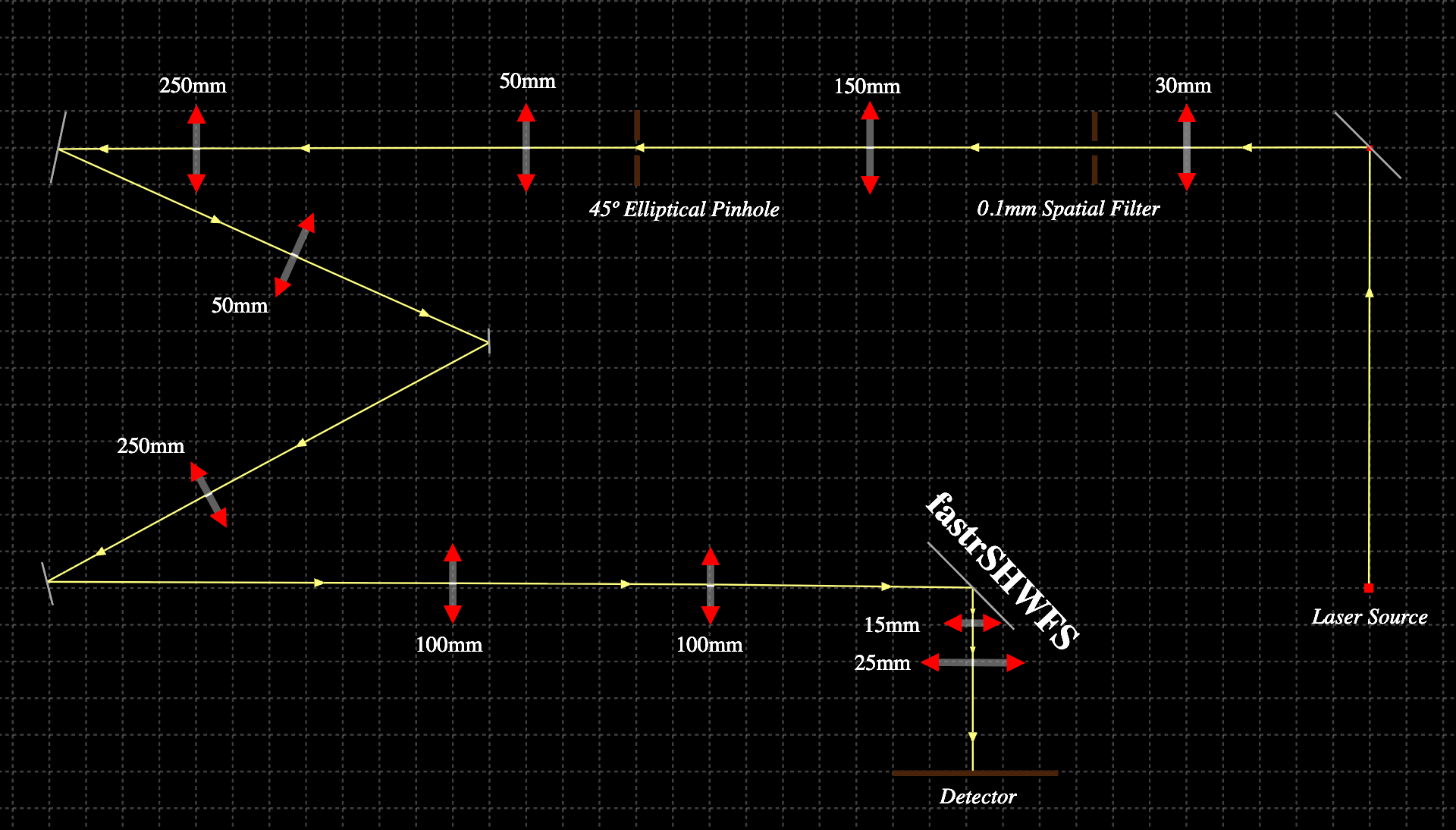}
   \end{tabular}
   \end{center}
   \caption[example] 
   { \label{fig:example} 
Diagram of laboratory layout for mask with focus.}
   \end{figure} 

The focus mask setup required lenses very close to the mask due to the high angle diverging rays that we saw in the plots earlier in this section. We ended up achieving close to telecentricity because the ½ inch lens could not be placed closer than $\sim 15mm$ away from the mask. And the 1-inch 25mm focal length lens was placed $\sim 25mm$ away from the ½ inch lens. These lenses are at a 90-degree angle from the upstream beam due to the reflection off the mask, elliptical pinhole geometry, and aligning the reflected beam with a column of holes in the testbed\footnote{See appendix A for beam reflection angle calculation}. The laboratory layout for the tip/tilt mask only differs downstream of the mask, with one 2-inch 60mm lens to focus spots onto the camera.

\subsection{Optical Quality}
The optical quality for both masks was examined using a Zygo white light interferometer which scanned the printed masks\footnote{Converting the Zygo files from ‘.dat’ to ‘.mat’ so they can be evaluated in Python was performed using a borrowed pre-made script.}. This plots are a subsection of each of the masks full cross-sections \footnote{See appendix B for full cross sections}.

\subsubsection{Focus Mask}
The focus mask mirrorlet subapertures were scanned around the middle of the mask. The cross sections in x and y were compared to the ideal design in Python, highlighting discrepancies in areas of the mask that were on the order of a micron shown in Figure 14. Good surface quality for optical elements have $\sim 50nm$ root mean square (RMS). Subtracting the design from the Zygo scan, we can see the residuals for the focus mask show a clear difference in design vs Zygo scan. Upon calculation, we achieved at best $\sim 1000nm$ RMS in a single subaperture. 

\subsubsection{Tip/tilt Mask}
The same analysis was performed for the tip/tilt mask, which yielded better results of $\sim 700nm$ RMS, but still not up to high quality standards shown in Figure 15. The central mirrorlet subaperture in the tip/tilt scan in the righthand plot can be ignored\footnote{The slope of the mask tilt in this subap is too steep for the Zygo scanner to read. The RMS analysis was therefore performed outside of this region. Appendix B has locations of cross sections wrt. the Zygo scan}. 

 \begin{figure} [H]
   \begin{center}
   \begin{tabular}{c} 
   \includegraphics[height=7cm]{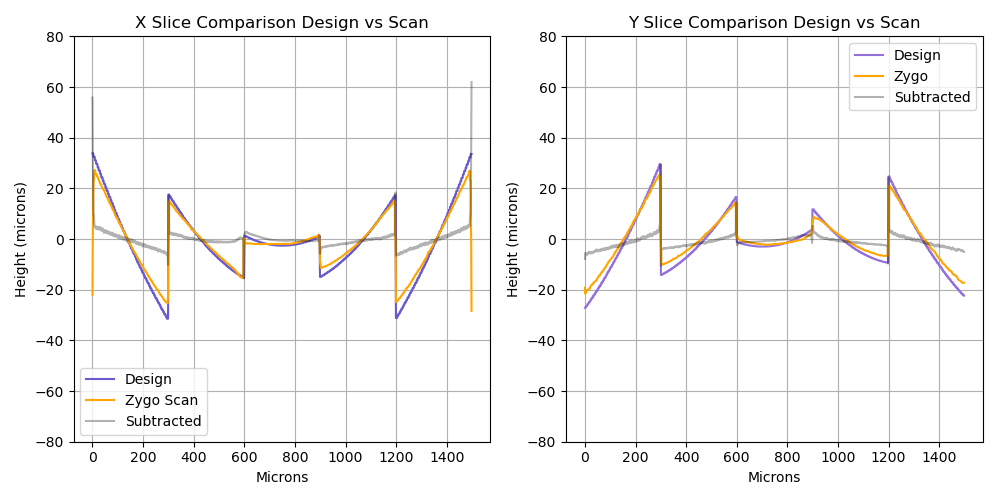}
   \end{tabular}
   \end{center}
   \caption[example] 
   { \label{fig:example} 
Central cross-section slices in x and y of focus mask scanned/printed mask in yellow, design in purple, and subtracted residual in grey. Axes in microns. }
   \end{figure} 

 \begin{figure} [H]
   \begin{center}
   \begin{tabular}{c} 
   \includegraphics[height=7cm]{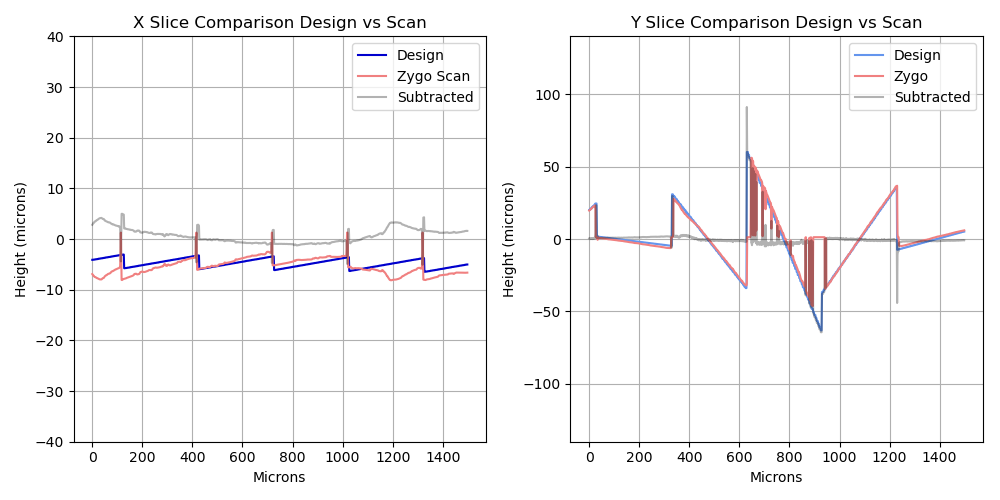}
   \end{tabular}
   \end{center}
   \caption[example] 
   { \label{fig:example} 
Central cross-section slices in x and y of tip/tilt mask scanned/printed mask in pink, design in blue, and subtracted residual in grey. Axes in microns. }
   \end{figure} 

\section{LAB RESULTS}
\label{sec:sections}

\subsection{Focus Mask}

The focus mask spot quality results are severely aberrated from lenses downstream of the mask imaged in Figure 16, imaged with the FLIR Blackfly Monochrome camera. The aberrations are most likely from rays at high angles entering the first lens at distances far from the principal axis. The high focal power of both lenses and the quality of the focusing in each mirrorlet implements aberrations that move the focal length for different rays away from the focal plane, meaning spots do not ever reach a line at the same z distance away from the mask. Shown in the following Figure 17, the spots at the focal length position does not converge to a line. 
 \begin{figure} [ht]
   \begin{center}
   \begin{tabular}{c} 
   \includegraphics[height=7cm]{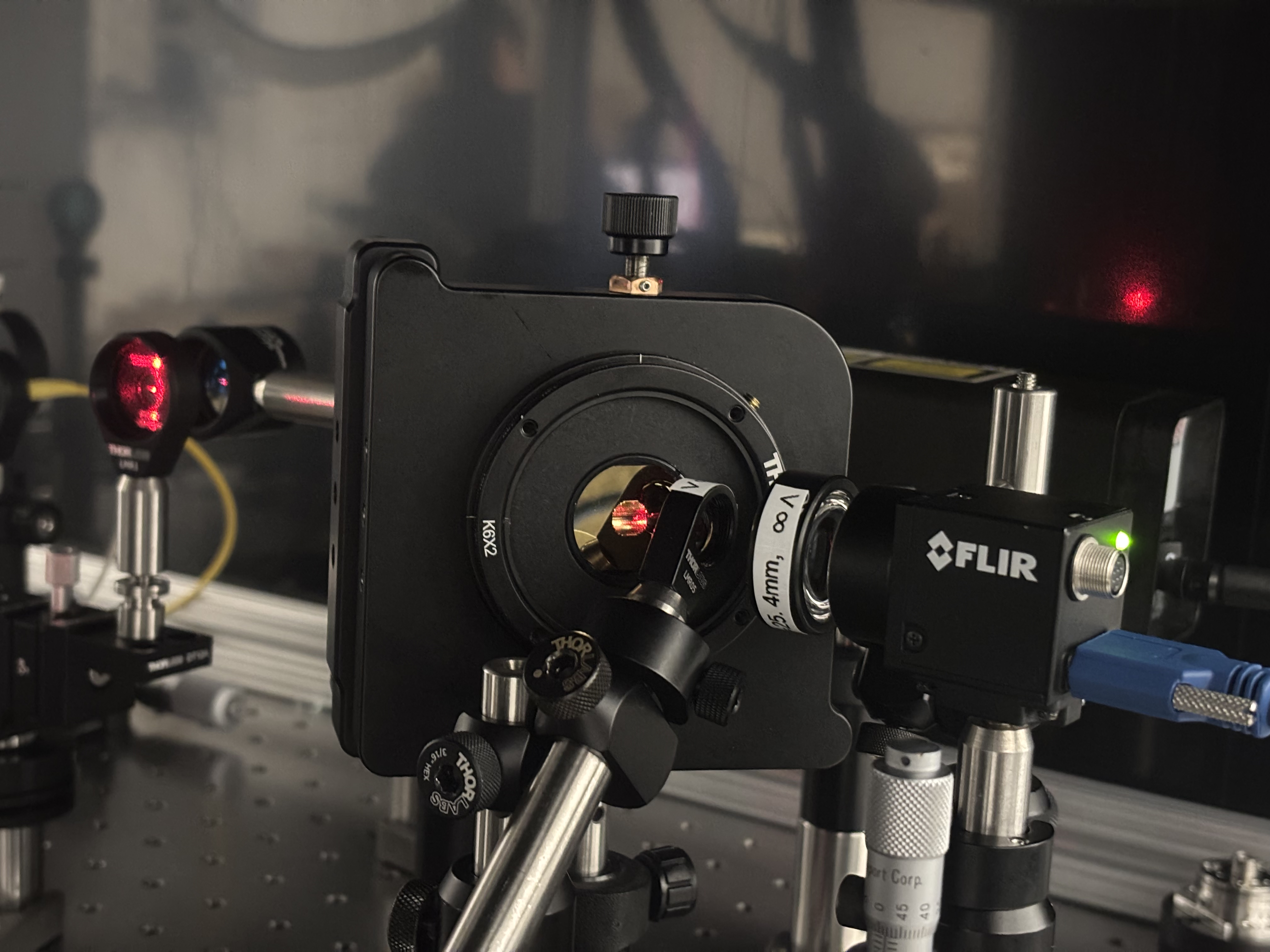}
   \end{tabular}
   \end{center}
   \caption[example] 
   { \label{fig:example} 
Photograph of the mask with focus, downstream lenses, and camera.}
   \end{figure} 
   
 \begin{figure} [ht]
   \begin{center}
   \begin{tabular}{c} 
   \includegraphics[height=4cm]{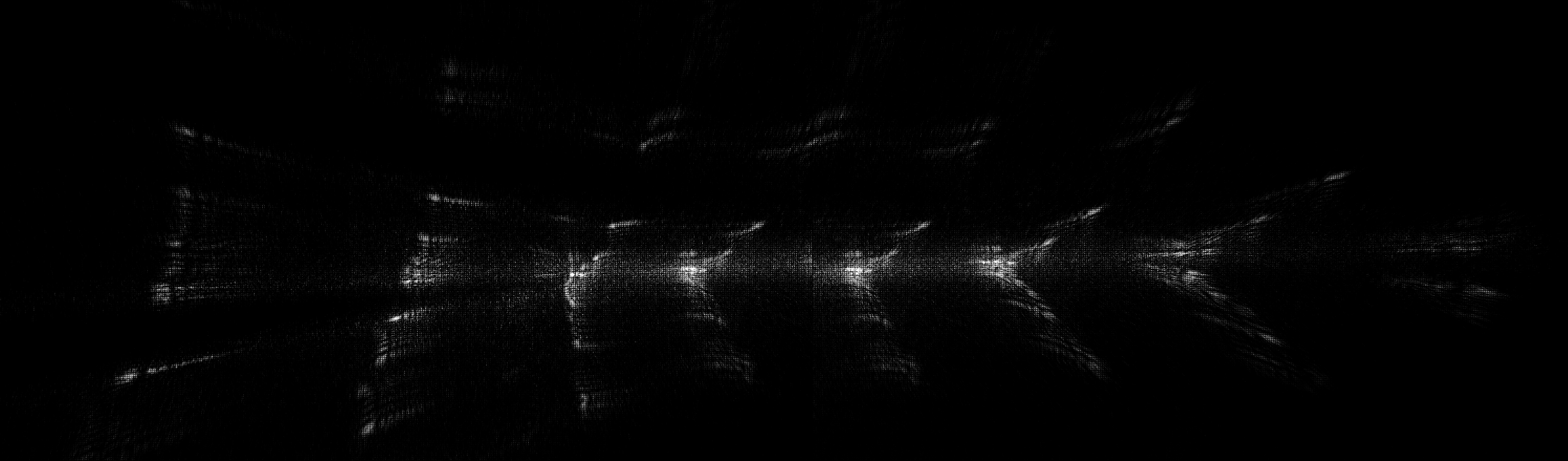}
   \end{tabular}
   \end{center}
   \caption[example] 
   { \label{fig:example} 
Best aligned focal plane image of focus mask, designed to redistribute spots into a line the width of the detector}
   \end{figure}

\subsection{Tip/tilt Mask}
The tip/tilt mask spot quality results are better, achieving a somewhat rectangular array throughout the middle but we can see effects of lens aberrations on both sides (compare to simulations in §3.2.2). The quality of the spots in the middle of the image (near the bright spot which is leakage of the beam around the mask) are due to the optical quality of the mask itself. While the quality of the spots at the ends, are due to rays entering the 2-inch lens at wide angles, aberrating their final position on the detector.

 \begin{figure} [ht]
   \begin{center}
   \begin{tabular}{c} 
   \includegraphics[height=1.7cm]{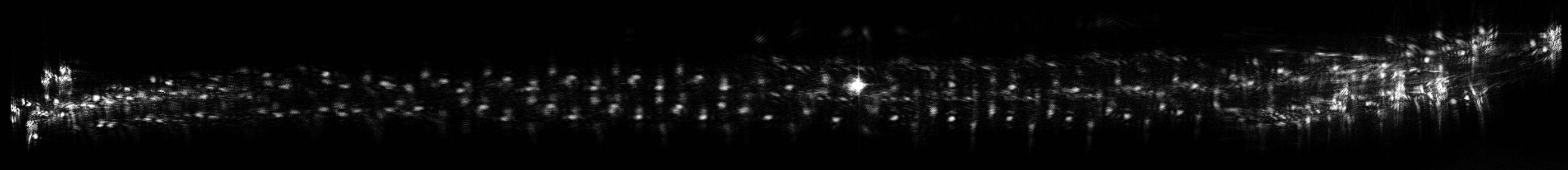}
   \end{tabular}
   \end{center}
   \caption[example] 
   { \label{fig:example} 
Best aligned focal plane image of tip/tilt mask, designed to redistribute spots into a 4 (or 5) x 100 rectangular array. }
   \end{figure} 

There may be a global tilt associated with the printing of the tip/tilt mask that yields a diagonal array of spots rather than distinct horizontal lines. Using a 300-micron pinhole on the tip tilt mask, we isolated the beam on one subaperture and tracing the beam in y direction along a strip of mirrorlets. This traced out a zig-zag sort of pattern on the detector in groups of three. This suggested bias associated with the mask printing itself in groups of three which we have yet to explore with Nanoscribe. For the best-defined region of spots to the left of the central axis in Figure 18, we propose a suspected orientation of the spot’s rows in Figure 19. The red circles should constitute the top row and so forth, each row originating from one of the vertical “strips” of the mask shown in the design in §3.1. The spots away from their respective axis are off by about a distance equal to 1.2 * FWHM from their expected row4. These diagonals look deceivingly like we placed our detector out of the focal plane so the process of how we confirmed focus is outlined in appendix C. One resolution element on the camera is $\sim 22.29$ pixels and our measured average FWHM is 26.46, meaning our spots are about 19\% larger\footnote{See appendix D for calculation} than the ideal diffraction‑limited resolution element. 

 \begin{figure} [ht]
   \begin{center}
   \begin{tabular}{c} 
   \includegraphics[height=4cm]{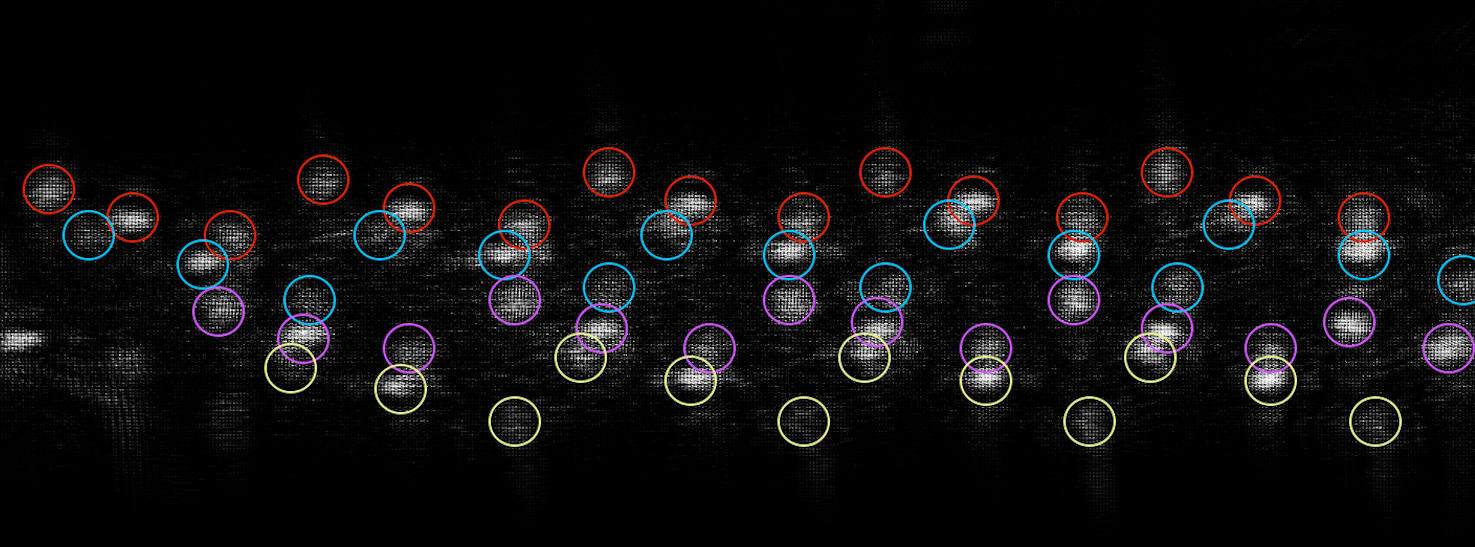}
   \end{tabular}
   \end{center}
   \caption[example] 
   { \label{fig:example} 
Best aligned focal plane image of tip/tilt mask, designed to redistribute spots into a 4 (or 5) x 100 rectangular array. }
   \end{figure} 

\section{CONCLUSION}
In this research essay, we outlined: §2 the standard AO system used in most ground-based telescopes and the current areas of limitations, §3 our solution to these limitations using the fastrSHWFS and its design, reflection geometry, optical layout and quality, and §4 test results for both masks using the LLNL AO testbed.  The reflection geometry yields high angle diverging rays, especially for the mask with focus, making the lab setup challenging and yielding severely aberrated results. The quality and reflection angle were slightly better for the tip/tilt only prototype, which is reflected in our results of a rectangular array of spots on the detector with some aberration. These masks are not up to standard according to surface quality, nor are they yielding good enough spot quality to perform the rest of an adaptive optics loop calculation. We would need to achieve diffraction limited spots with minimal aberration in order to use the quad-cell slope reconstruction that the current SHWFS uses to correct for Earth’s atmosphere. A mask with tip/tilt only with good surface quality is a more promising route for this technique, and a new version would hopefully yield better results for a respective lab setup. 
\vspace{2cm}
\acknowledgments 
 
This work was performed under the auspices of the U.S. Department of Energy by Lawrence Livermore National Laboratory under Contract DE-AC52-07NA27344. This document number is LLNL-PROC-2017376. This work was also supported in part by the U.S. Department of Energy, Office of Science, Office of Workforce Development for Teachers and Scientists (WDTS) under the Science Undergraduate Laboratory Internships (SULI) Program. Thank you to my supervisor, Mark Ammons, and to members of the adaptive optics group, including Bautista Fernandez, Cesar Laguna, and Mike Kim, for their help in the lab. I would also like to thank Gary Tham for his assistance with the Zygo optical scanner. 

\appendix

\section{Incoming Beam Modeling \& Lab Setup}
\label{appendix:A}
We were concerned that with the reflective property of the focus mask, we would obstruct the incoming beam with a 1- or 2-inch lens placed downstream. We calculated the minimum angle at which the mask needed to be placed to not obstruct the incoming beam, requiring a ½ inch lens shown in Figure 20. This is why in our laboratory setup, we chose a high powered ½ inch lens, even though our setup eventually had 90º reflection, meaning a 1-inch lens would not have obstructed the beam.

 \begin{figure} [ht]
   \begin{center}
   \begin{tabular}{c} 
   \includegraphics[height=8cm]{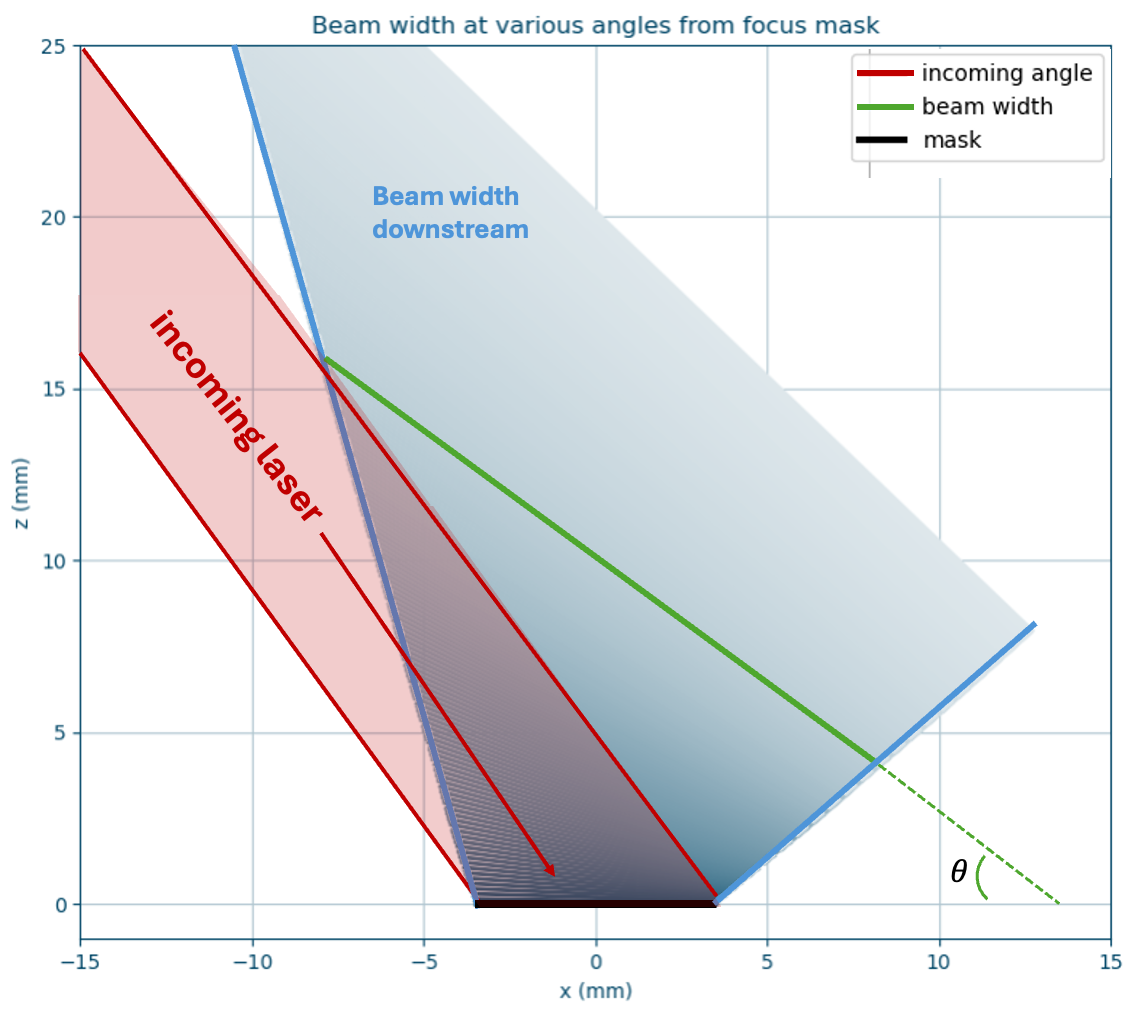}
   \end{tabular}
   \end{center}
   \caption[example] 
   { \label{fig:example} 
Minimum reflection angle for an incoming beam angle of $\sim 55$ degrees (red). Shows the mask’s reflected beam angle in green for that incoming angle. Beam width is shown by the green line. A 1 inch lens placed at an angle theta would not obstruct the incoming beam for this given laser angle.}
   \end{figure} 
We used a 0.1mm spatial filter to remove high spatial frequencies from the laser and make our beam a top-hat function with a hard-edge pupil. We used a 45º elliptical pinhole to achieve a circular beam on the mask which is at a 45º angle to the oncoming beam. The 50mm and 250mm pair lenses are for resizing the beam width from 0.48mm to $\sim 7 mm$ across to fully illuminate the mask size which is 6.9mm across. The 100mm pair are reimaging lenses to move the pupil downstream to place the mask in a more accessible place on the testbed.

 \begin{figure} [H]
   \begin{center}
   \begin{tabular}{c} 
   \includegraphics[height=5cm]{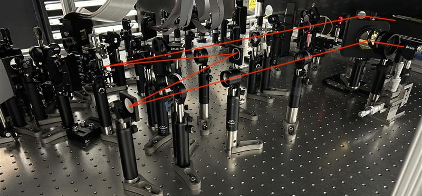}
   \end{tabular}
   \end{center}
   \caption[example] 
   { \label{fig:example} 
Lab setup for mask with tip/tilt only}
   \end{figure} 
A photograph of the laboratory setup for the tip tilt mask is shown in Figure 21. The only change that is made compared to the mask with focus was swapping out the mask in the mount and replacing the two lenses with one 2 inch lens shown on the righthand side of the image.

\section{Cross Sections of Masks}
\label{appendix:B}
The cross sections for the entire width of both masks in x and y directions are shown in Figure 21. The Zygo optical quality scan only shows $\sim 5$ mirrorlets, which is a small section of these full scans, that section is shown in Figure 22. The discrepancies between the design and the printed result for either mask could be caused by: 1) the .dat to .mat conversion script implementing false conversions into our file from pre-made assumptions, 2) the Zygo scanning could have inaccuracies in properly surveying the surface of steep elements of both masks, and/or 3) could be due to an actual optical fault in the printing of the masks themselves. In the next section, we outline the test results of both masks.  

 \begin{figure} [H]
   \begin{center}
   \begin{tabular}{c} 
   \includegraphics[height=6cm]{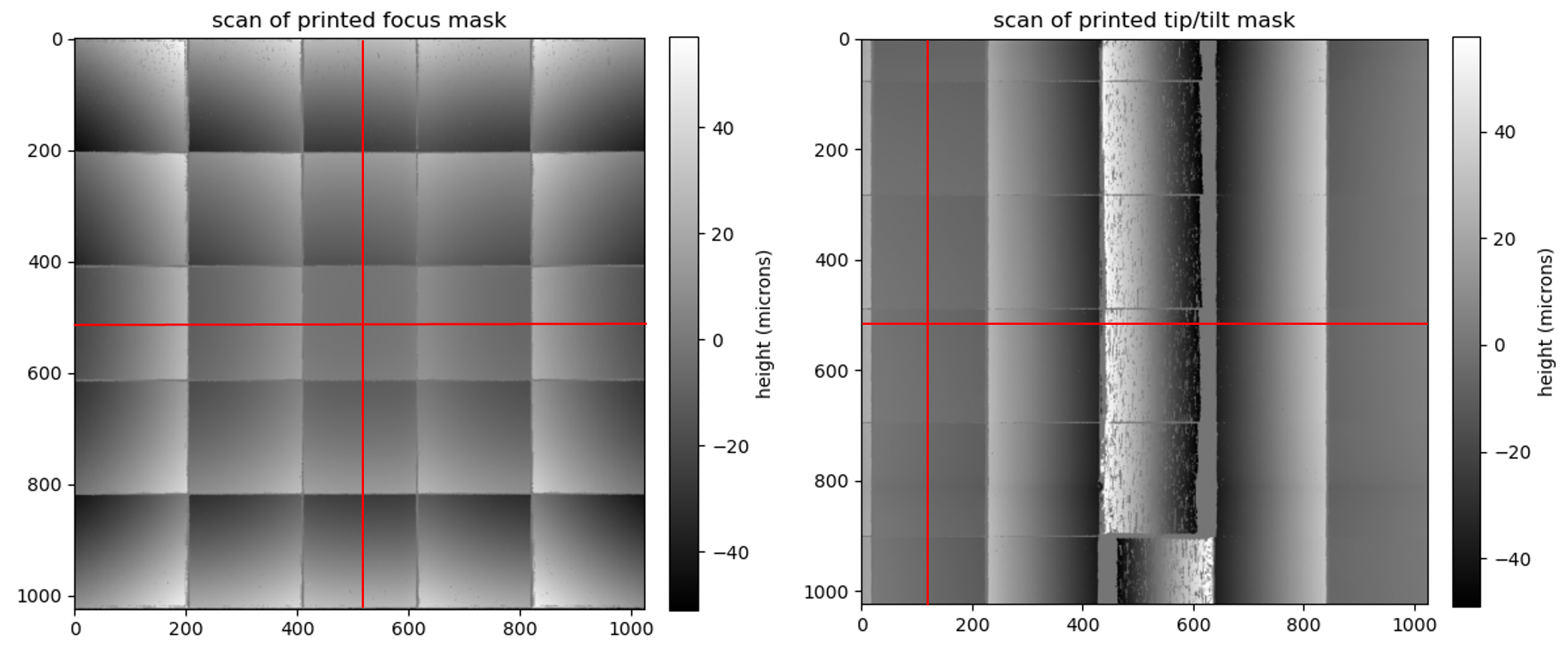}
   \end{tabular}
   \end{center}
   \caption[example] 
   { \label{fig:example} 
Zygo scans of a $\sim 5 \text{x}5$ subaperture region of both printed masks. Red lines indicate where RMS and cross sections were plotted in 3.4}
   \end{figure} 

 \begin{figure} [H]
   \begin{center}
   \begin{tabular}{c} 
   \includegraphics[height=9cm]{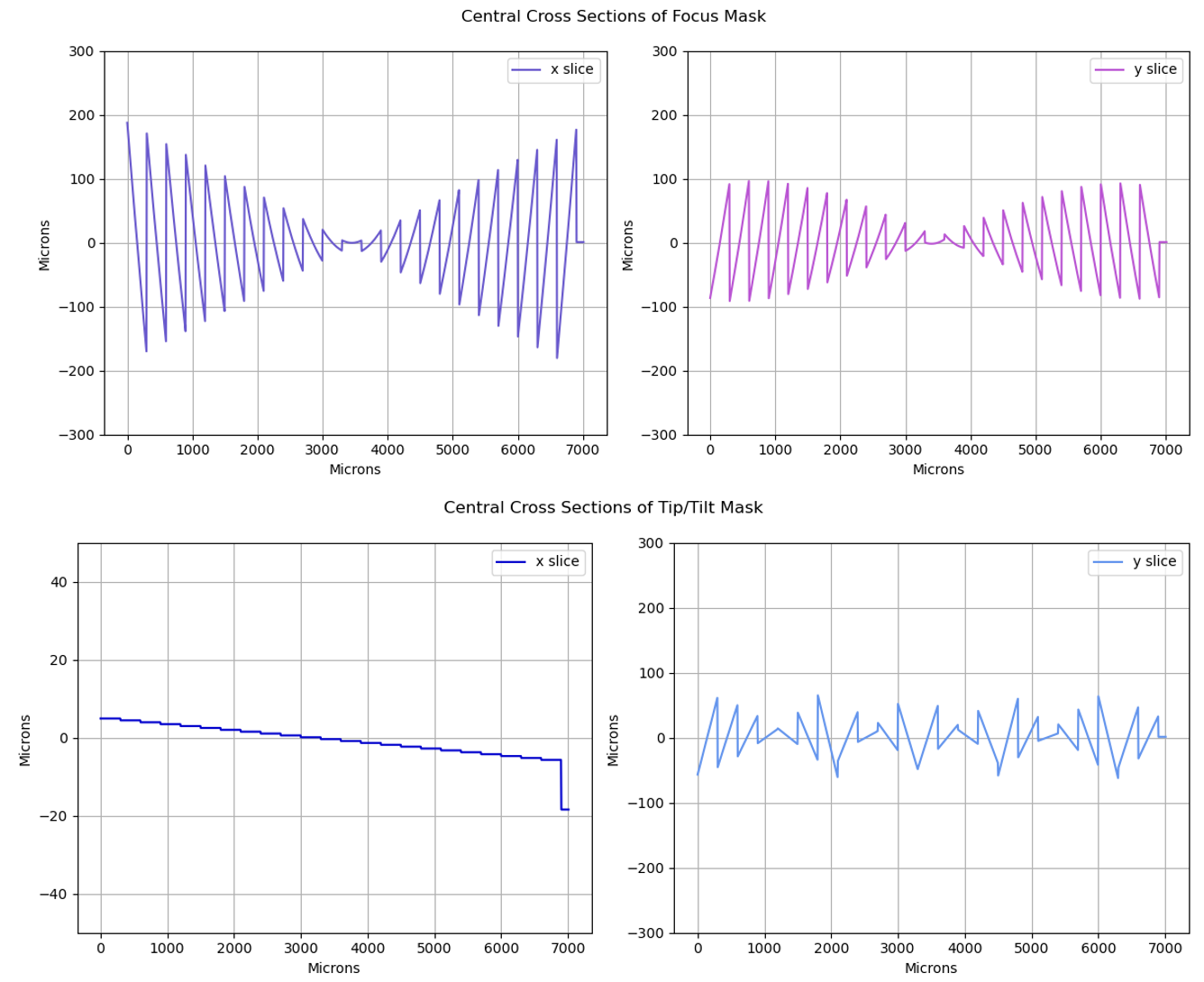}
   \end{tabular}
   \end{center}
   \caption[example] 
   { \label{fig:example} 
Entire cross sections of focus mask (top) and tip/tilt only mask (bottom)}
   \end{figure} 

\section{Focal Plane Confirmation}
\label{appendix:C}

For the mask with focus, there is no pre-calibrated focal plane, so we must estimate where the focal plane lies downstream of the lenses. Figure 24 shows what the spots look like as you approach the focal plane for the mask with focus a few millimeters upstream of the focal plane. This shows the nature of the aberration due to the lenses. A 300-micron pinhole was used for both optical test setups. This was used to illuminate one mirrorlet at a time in either mask to see individual spot quality without possible interference between mirrorlets. An attempt to illuminate only one mirrorlet for the focus mask is shown in Figure 25. Obviously, neighboring mirrorlets have been illuminated, resulting in 3 spots on the detector. This may be due to beam size growing slightly over 300 microns in our lab setup which is very possible.
    \begin{figure} [ht]
    \begin{center}
    \begin{tabular}{c} 
    \includegraphics[height=2.5cm]{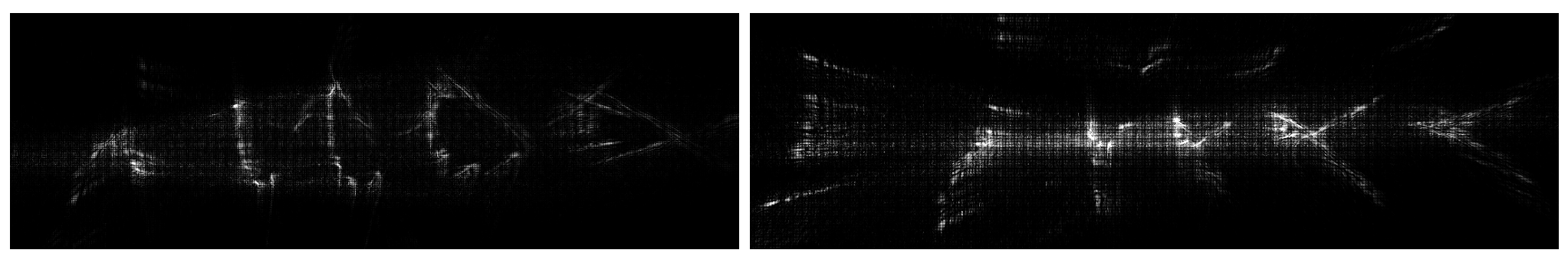}
    \end{tabular}
    \end{center}
    \caption[example] 
    { \label{fig:example} 
    Focus mask spots a few mm upstream of estimated focal plane}
    \end{figure} 
   
    \begin{figure} [ht]
    \begin{center}
    \begin{tabular}{c} 
    \includegraphics[height=2.5cm]{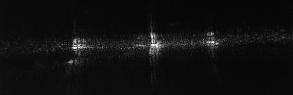}
    \includegraphics[height=2.5cm]{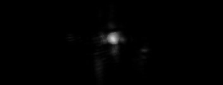}
    \end{tabular}
    \end{center}
    \caption[example] 
    { \label{fig:example} 
    Attempt at illuminating one subaperture in the focus mask (left) and tip/tilt mask (right}
    \end{figure} 

The focus mask has a pre-calibrated focal plane to help us ensure we are placing the detector at focus. Placing both the detector and the mask on linear stages, we were able to move the mask back, meaning the beam hit the reflective (can be thought of merely as a fold mirror) portion of the mask rather than the inscribed portion. This reflected beam was used to align both lenses downstream. The beam saturated the detector, so we placed a 20A ND filters into a collimated space of the optical setup and a 400-micron pinhole in the pupil plane upstream of the mask. Moving the camera on the linear stage, we were able to zoom in on the unsaturated illuminated area of the detector until it came to focus. After focus was found, we moved the mask back onto the inscribed portion. Knowing the camera was at focus, the images were then taken and stitched together for the tip/tilt mask shown earlier in Figure 18. We were able to isolate the mirrorlet slightly better for the mask with tip/tilt only in Figure 25 on the right.

\section{Resolution Element Calculation}
\label{appendix:D}
1 Resolution element calculated using: 60mm focal length lens, 300 $\mu$m mirrorlet, 633 nm light, 5.68 microns per pixel (plate scale). Fitting a Gaussian to a standard spot (similar to the size in Figure 24 for the tip tilt mask) using a centroiding algorithm yields a measured FWHM size of 26.46 pixels using the SciPy package in Python. 

\begin{equation}
\frac{60}{0.3} \times \frac{0.000633}{0.00568} = 22.29 \ \text{pixels}
\end{equation}

\begin{equation}
\frac{\text{Measured}}{\text{Expected}} = \frac{26.46}{22.29} = 1.19
\end{equation}

\begin{equation}
(1.19 - 1.0) \times 100 = 19\% \ \text{larger}
\end{equation}

\vspace{2cm}

\bibliographystyle{spiebib} 

\end{document}